\documentclass[twocolumn,showpacs,preprintnumbers,amsmath,amssymb]{revtex4}

\usepackage{graphicx}
\usepackage{dcolumn}
\usepackage{bm}

\begin{document}

\title{Anomalous dissipation in the mixed state of underdoped cuprates
close to the superconductor-insulator boundary}

\author{C. Capan$^\dag$, K. Behnia} \affiliation{Laboratoire de Physique Quantique,
 (UPR 5 CNRS), ESPCI, 75231 Paris, France}
\author{Z. Z. Li, H. Raffy}\affiliation{Laboratoire de
Physique des Solides (UMR 8502 CNRS), Universit\'e Paris-Sud,
91405 Orsay, France},
\author{C. Marin}\affiliation{DRFMC/SPSMS, Commisariat \`{a}
l'Energie Atomique, 38042 Grenoble, France}

\date{December 17, 2002}

\begin{abstract}
We present a comparative study of Nernst effect and resistivity in
underdoped samples of Bi$_2$Sr$_2$CuO$_{6+\delta}$ and
La$_{2-x}$Sr$_{x}$CuO$_4$. The Nernst effect presents a peak in a
region of the H-T diagram where resistivity shows a non-metallic
temperature dependence. Our results illustrate that the mechanism
of dissipation in the mixed state of underdoped cuprates is poorly
understood. Large quantum superconducting fluctuations and
vanishing vortex viscosity are among suggested explanations for an
enhanced Nernst signal close to the superconductor-insulator
boundary.

\end{abstract}

\maketitle

Recently, the peculiar behavior of Nernst effect in copper oxide
superconductors has become a subject of growing
attention\cite{ong1,ong2,ong3,capan,sondhi,kontani,weng,ikeda,ikedater}.
In conventional superconductors, Nernst effect, namely the
transverse component of the thermopower in a magnetic field, is
known to be associated with vortex movement\cite{huebenerbook}.
However, Xu \emph{et al.}\cite{ong1} reported that in  underdoped
La$_{2-x}$Sr$_x$CuO$_4$ a large residual signal persists in the
normal state and well above T$_c$. This observation has been
confirmed in other families of cuprate superconductors including
Bi$_2$Sr$_2$CuO$_{6+\delta}$ and
YBa$_2$Cu$_3$O$_{6+y}$\cite{ong2,ong3}. The normal-state Nernst
signal is present over a wide doping range, but attenuates as one
moves towards the superconductor-insulator boundary or the
overdoped regime in the phase diagram\cite{ong2}. Remarkably, the
temperature scale up to which such an anomalous Nernst signal
extends is about 130K, regardless of the particular system
studied. Moreover, the Nernst signal is found to persist at
magnetic fields as high as 30T both in overdoped\cite{ong3} and in
underdoped\cite{capan} compounds.

The debate on the origin of this residual Nernst signal in the
normal state of underdoped cuprates is still open and has
stimulated a number of theoretical works during the past months.
Fluctuations of the superconducting order parameter above T$_c$
are considered to be the most plausible explanation for such a
signal. Ussishkin, Sondhi and Huse\cite{sondhi}, for example, used
a Time-Dependent-Ginzburg-Landau(TDGL) approach. Within gaussian
approximation and subtracting magnetization currents, they found a
sizeable Nernst signal due to superconducting fluctuations. Weng
and Muthukumar\cite{weng} have suggested that in a
Resonant-Valence-Bond(RVB) picture, the coupling of spinon
vortices to holons, can give rise to an enhanced Nernst effect
above T$_c$ in the so-called spontaneous vortex phase. A different
route has been taken by Kontani\cite{kontani} who calculated
Nernst coefficient beyond the Relaxation Time Approximation. By
taking a self-consistent account of vertex corrections for
currents, he found that the quasi-particle contribution to Nernst
effect is no longer negligible in presence of antiferromagnetic
and superconducting fluctuations. These studies point to the
presence of  strong superconducting fluctuations in the pseudogap
regime and  suggest that the superconducting transition is not a
mean-field transition\cite{levin} but rather a vortex-antivortex
unbinding type of transition\cite{tesanovic}. It is puzzling,
however, that these strong fluctuations are not detected in charge
transport. Indeed, in the temperature window associated with the
residual Nernst signal, the magnetoresistance do not present any
strong feature.

In this paper, we focus on the behavior of the Nernst effect in
the underdoped regime close to the supercoductor-insulator
transition. In underdoped cuprates, the destruction of
superconductivity in high magnetic fields is accompanied with the
emergence of an ``insulating" normal state presenting a weakly
diverging non-metallic resistivity\cite{ando,boebinger,ono}. Our
study concentrates on very underdoped samples of
La$_{2-x}$Sr$_x$CuO$_4$ and Bi$_2$Sr$_2$CuO$_{6+\delta}$ close
enough to the superconductor-insulator boundary for a moderate
magnetic field of 12T  to introduce a non-metallic resistivity. By
simultaneously measuring Nernst effect and resistivity in the
mixed state, we found that these two transport properties, both
supposedly associated with vortex motion, are no longer correlated
in this regime. In sharp contrast with other superconductors, the
maximum in the Nernst occurs in a temperature window associated
with a non-metallic resistivity. This result highlights the limits
of our current understanding of the dissipation mechanism in a
field-induced superconductor-insulator transition.

Nernst coefficient, the ratio of the transverse electric field to
the longitudinal thermal gradient, was measured in presence of a
magnetic field  parallel to the c-axis with a
one-heater-two-thermometer set-up. The set-up allowed to measure
the resistance of the sample in the same conditions. The thermal
gradient was obtained using two RuO$_2$ or Cernox thermometers
connected via gold or silver wires to two electrodes along the
sample. The same electrodes were used to measure the voltage drop
induced by a charge current applied along the sample and therefore
to monitor the resistance of the sample. Two other lateral
electrodes were used to measure the DC transverse voltage produced
by applying a heat current along the sample with a heater chip.
The two thermometers and the heater were held in the vacuum by
long thin highly resistive manganin wires which were also used to
measure their resistance. In this way, the path of the heat
current along the sample was controlled. No correction was made
for the magneto-resistance of thermometers. We estimate that the
error on the absolute magnitude of the temperature gradient at 12T
due to this approximation is less than 10 percent. Note that such
a correction is only relevant for single crystals; for thin films,
the thermal gradient is almost totally controlled by the
substrate's thermal conductivity and is not expected to vary with
the magnetic field. The heat current was kept low enough to keep
$\nabla$T/T below 5$\%$. At each temperature, the Nernst signal
was extracted by reversing the sign of the magnetic field and
keeping only the antisymmetric part of the signal. In this way,
the offset signal due to the misalignment of the contacts and an
eventual contribution of the wires was subtracted.

The samples used in this study were two c-axis oriented thin films
of Bi$_2$Sr$_2$CuO$_{6+\delta}$(Bi-2201) and a single crystal of
La$_{1.94}$Sr$_{0.06}$CuO$_4$. The Bi-2201 thin films were
epitaxially grown by RF magneton sputtering on a SrTiO$_3$
substrate\cite{bi2201}. The oxygen content of initially overdoped
samples was reduced by annealing in controlled atmosphere as
already detailed in a previous study of the evolution of
resistivity with doping in
Bi$_2$Sr$_{2-z}$La$_z$CuO$_{6+\delta}$\cite{kons}. Gold electrodes
were sputtered on them and wires were connected with silver paint
to these electrodes. The La$_{1.94}$Sr$_{0.06}$CuO$_4$ single
crystal was made in an optical furnace using a travelling-solvent
floating zone technique and was already used and mentioned in our
previous study\cite{capan}. It was subsequently oxygen annealed at
10 bar and 450C for 10 days in order to enhance the homogeneity of
oxygen distribution. Low-resistivity contacts on this crystal were
made by baking Dupont 6838 silver paint electrodes with oxygen
flow at 450C for 10 minutes.

\begin{figure}
\resizebox{!}{0.7\textwidth}{\includegraphics{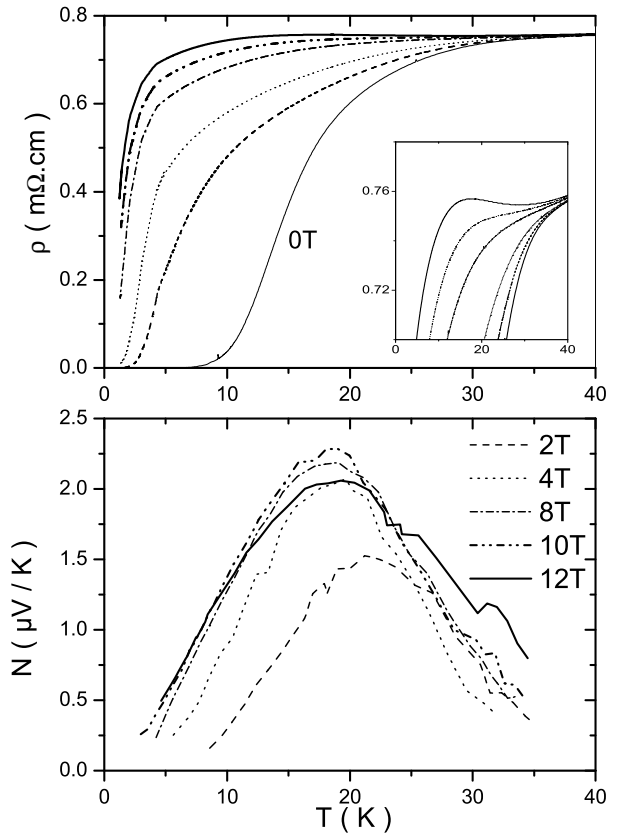}}
  \caption{\label{BiaRhoN}Nernst effect(bottom) and resistivity(top) as a function of temperature up
  to 12T in Bi$_2$Sr$_2$CuO$_{6+y}$ thin film with T$_c$=15K. The insert
  highlights the field-induced upturn in resistivity.}
\end{figure}

\begin{figure}
\resizebox{!}{0.7\textwidth}{\includegraphics{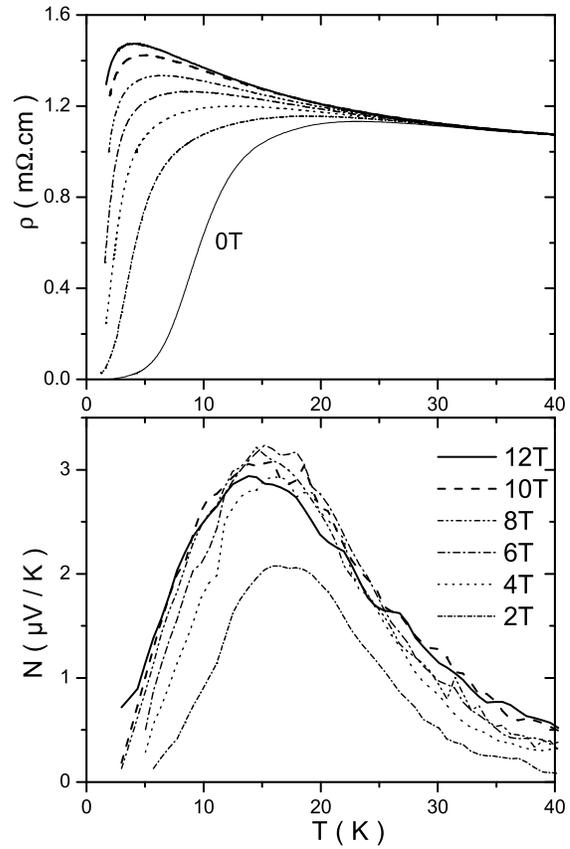}}
  \caption{\label{BibRhoN} Nernst effect(bottom) and resistivity(top)
  as a function of temperature up to 12T in Bi$_2$Sr$_2$CuO$_{6+y}$ thin
  film with T$_c$=9K.}
\end{figure}

Figures ~\ref{BiaRhoN} and ~\ref{BibRhoN} show Nernst effect and
resistivity as a function of temperature in two
Bi$_2$Sr$_2$CuO$_{6+\delta}$ thin films, at several magnetic
fields up to 12T. At zero field, both samples present a broad
resistive transition which shifts to lower temperatures with
increasing magnetic field. The critical temperature, taken as the
temperature associated with resistivity dropping to half of the
full normal-state value, was found to be 15K for the first and 9K
for the second sample. The resistivity of the 9K sample presents a
clear non-metallic upturn which becomes more pronounced as the
field is increased. In the case of 15K sample, the non-metallic
behavior is absent at low fields but sets in at 12T, indicative of
a higher doping level. Using the critical temperature and the
room-temperature resistivity of these samples, one can make a very
rough estimate of doping level by comparing them with values
obtained in systematic studies of doping dependence of
resistivity\cite{kons,rifi} and thermopower\cite{kons2}. This
yields $\delta\sim$ 0.08 for the 9K sample and $\delta\sim$ 0.09
for the 15K sample.

In both samples, Nernst effect presents the broad peak commonly
associated with the vortex movement in the mixed state of type II
superconductors.  The amplitude of this peak increases initially
with increasing magnetic field before starting to decrease at 10T.
But the peak \emph{does not} shift to lower temperatures as the
field is increased. This is in sharp contrast with the behavior
observed in optimally-doped cuprates, where the Nernst peak
follows the field-induced shift in resistive
transition\cite{huebener}. Note that in both samples at 12T, the
peak occurs at a temperature interval where resistivity is
increasing with decreasing temperature. In the case of 15K sample,
while resistivity presents a continuous evolution from the
low-field metallic to the high-field localization behavior, the
temperature dependence of the Nernst effect remains unaffected.
This, added to the fact that Nernst effect has a similar behavior
in both samples, seems to suggest that that Nernst effect is not
disturbed by the onset of localization observed in resistivity.
Comparing the two Bi-2201 thin films, one important feature is
that the mismatch between the Nernst peak and the resistive
transition becomes more pronounced as the magnetic field is
increased or the doping is reduced at constant field. As these are
the parameters that tune the superconductor-insulator transition,
the anomalous dissipation that we report here might be related to
this transition. Note, however, that the threshold field for
non-metallic resistivity does not correspond to the onset of this
absence of correlation between Nernst effect and resistivity.

\begin{figure}
\resizebox{!}{0.7\textwidth}{\includegraphics{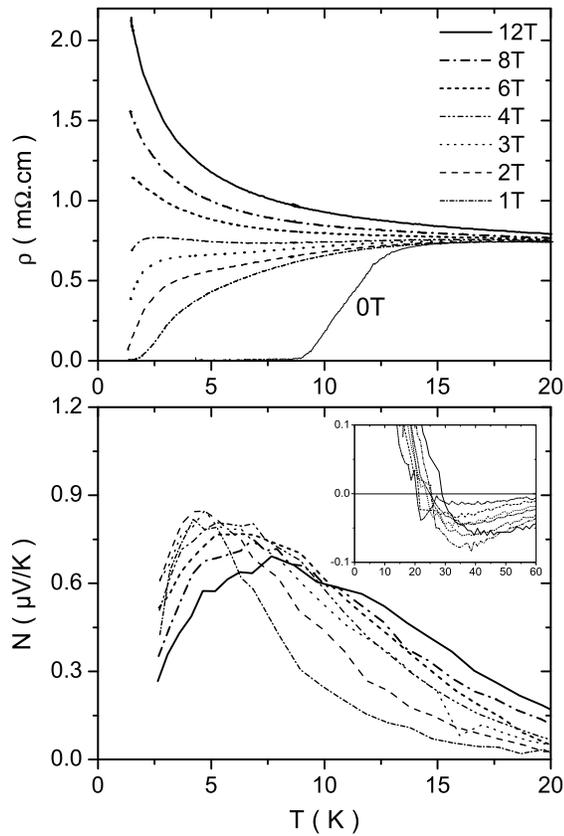}}
  \caption{\label{Lsco06RhoN}Nernst effect(bottom) and resistivity(top) as a function of temperature up
  to 12T in La$_{1.94}$Sr$_{0.06}$CuO$_4$ single crystal with T$_c$=11.4K.
  The insert shows the sign change of the Nernst signal at higher temperatures.}
\end{figure}

This contrasting behavior of Nernst effect and resistivity is even
more pronounced in the La$_{1.94}$Sr$_{0.06}$CuO$_4$ as seen in
figure ~\ref{Lsco06RhoN}. As a result of an extensive oxygen
annealing, this sample presents a smooth zero-field resistive
transition centered around 11.4K in contrast with the broad double
transition seen before annealing\cite{capan}. Nevertheless, the
contrasting behavior of Nernst effect and resistivity was found to
be robust to oxygen annealing and this rules out extrinsic
inhomogeneity as a primary cause for it. As seen in the figure,
for fields larger than 6T, the resistivity does not show any sign
of superconducting transition down to 1.2K. Note, however, that,
even taken alone, this behavior of resistivity does not indicate a
complete destruction of superconductivity, since a positive
magnetoresistance is present up to 12T. In the same temperature
range, we observe a peak in Nernst effect which shifts to higher
temperature with increasing magnetic field. After an initial
increase, the amplitude of this peak begins to decrease with
increasing magnetic field for fields exceeding 2T. Moreover, as
seen in the insert, the Nernst signal changes sign well above
T$_c$, revealing a small quasi-particle contribution with opposite
sign to the vortex Nernst signal, consistent with previous
reports\cite{ong2}.

Across a wide range of the cuprate phase diagram, the Nernst
effect and resistivity do not give identical accounts of the way
magnetic field destroys superconductivity. Recent high-field
investigations of Nernst effect in overdoped
La$_{1.8}$Sr$_{0.2}$CuO$_4$ indicate that the upper critical field
extracted from the resistivity measurements is systematically
lower than the H$_{c2}$ obtained from the Nernst
signal\cite{ong3}. A similar discrepancy  has also been reported
in the case of electron-doped Nd$_{2-x}$Ce$_x$CuO$_4$ at different
doping levels\cite{gollnik}. Our results indicate that the
decoupling between the two probes of vortex state becomes
particularly striking by approaching to the insulating side of the
phase diagram and in the presence of field-induced non-metallic
resistivity.

Recently, in order to explain the absence of correlation between
Nernst effect and resistivity, Ikeda proposed a model based on
quantum superconducting fluctuations described by a 2D
Ginzburg-Landau action\cite{ikeda}. Noticing that in underdoped
cuprates, the normal-state resistance per CuO$_2$ plane becomes
comparable to the quantum of resistance
R$_q$=$\frac{h}{4e^2}$=6.45k$\Omega$, he argued that in the
underdoped regime, as a result of a reduced superfluid density,
superconducting fluctuations are not only enhanced, but become
quantum in nature in contrast to the optimally-doped regime
dominated by thermal fluctuations. The same formalism has already
been used to describe quantum fluctuations in the context of the
field-tuned superconductor-insulator transition in dirty
two-dimensional superconducting films\cite{goldman}, accounting
for non-universal values of resistance and the absence of a
well-defined critical field at the transition by including effects
of Coulomb repulsion\cite{ikedabis}.

In Ikeda's scenario\cite{ikeda}, the apparent contradiction
between a vortex-like behavior in Nernst effect and a normal-state
like resistivity is due to a vanishingly small vortex contribution
to the total resistivity which is thus dominated by the insulating
background, whereas the Nernst effect is enhanced in the quantum
fluctuation regime. Decomposing the pair wavefunction on Landau
levels and taking into account the $\omega\neq0$ terms, the theory
provides expressions for transport entropy and conductivity in
presence of quantum fluctuations as a function of the propagator
of the first two Landau level fluctuation fields. This model can
mimic the observed behavior of both transport energy and
resistivity reported here in the high temperature window above the
Nernst peak\cite{ikedater}.

The doping dependence of vortex viscosity may constitute an
alternative explanation for the observed discrepancy between
resistivity and Nernst data. The vortex equation of movement
contains a dissipative term proportional to vortex viscosity
$\eta$\cite{huebenerbook}. Since the Nernst signal is inversely
proportional to this viscosity (N$\sim \frac{S_{\Phi}}{\eta}$,
with S$_{\Phi}$ the vortex entropy), a vanishing viscosity would
lead to an enhanced Nernst signal as well as a rapid saturation of
resistivity to its normal state value. A scenario along these
lines was suggested to account for the anomalous behavior of the
resistive transition in overdoped
Tl$_2$Ba$_2$CuO$_{6+\delta}$\cite{geshkenbein}. In the latter
compound, the apparent H$_{c2}$ extracted from resistivity lie
below the thermodynamic one deduced from specific heat
data\cite{mackenzie}. Experimental evidence for a reduced vortex
viscosity in overdoped Bi-2201 has been recently provided by
microwave surface impedance measurements at low
fields\cite{matsuda}. On the other hand, studying the properties
of a d-wave superconductor in a slave boson-U(1) gauge theory,
Ioffe and Millis proposed that viscosity should vanish in the
vicinity of the Mott insulator\cite{millis}. Therefore, in such a
picture, the coexistence between an enhanced Nernst signal and a
saturated flux-flow resistivity is expected to enhance with
underdoping. Further investigations of the doping dependence of
vortex viscosity are necessary to shed more light on this issue.

Finally, we note that recent Scanning Tunnelling microscopy
studies indicate that vortex cores in cuprates present an unusual
electronic excitation spectrum\cite{hoffman}. Moreover, in the
underdoped regime, inelastic neutron scattering\cite{lake} and
Nuclear Magnetic Resonance\cite{mitrovic,mitrovic2} experiments
have revealed enhanced antiferromagnetic correlations associated
with vortex cores. The possible contribution of these excitations
to the Nernst signal is yet to be explored.

In conclusion, we found that the decoupling between Nernst effect
and resistivity becomes more pronounced as one sweeps the cuprate
phase diagram from optimal doping to the very underdoped limit. In
particular, the Nernst effect presents a peak where the
corresponding resistivity is non-metallic.  Such an anomalous
dissipation regime may prove to be relevant for the studies of
superconductor-insulator transition in dirty conventional
superconducting thin-films.

We acknowledge helpful discussions with R. Ikeda, H. Kontani and
I. Ussishkin.

$^\dag$\emph{Present address: Los Alamos National Laboratory, Los
Alamos, NM 87545}

\end{document}